\documentclass{elsart}
\usepackage{graphicx,amssymb}
\newif\ifpdf\ifx\pdfoutput\undefined\pdffalse\else\pdfoutput=1\pdftrue\fi

\journal{NIM A}
\begin{document}
\begin{frontmatter}
\title{Study of the Acoustic Signature of UHE Neutrino Interactions in Water and Ice}
\author[acorne]{The ACORNE collaboration:},
\author[UCL]{S.~Bevan},
\author[Sheffield]{A.~Brown}\footnote{Now at C.P.P.M. Marseille.}  
\author[Northumbria]{S.~Danaher\corauthref{cor}},
\corauth[cor]{Author for correspondence}
\ead{sean.danaher@unn.ac.uk}
\author[Sheffield]{J.~Perkin},
\author[IC]{C.~Rhodes},
\author[Lancaster]{T.~Sloan},
\author[Sheffield]{L.~Thompson},
\author[Sheffield]{O.~Veledar},
\author[UCL]{D.~Waters}
\address[acorne]{\texttt{http://pppa.group.shef.ac.uk/acorne.php}}
\address[UCL]{Department of Physics \& Astronomy, University College London, U.K.}
\address[Northumbria]{CEIS, University of Northumbria, U.K.}
\address[Sheffield]{Department of Physics \& Astronomy, University of Sheffield, U.K.}
\address[IC]{Institute for Mathematical Sciences, Imperial College London, U.K.}
\address[Lancaster]{Department of Physics, University of Lancaster, U.K.}

\begin{abstract}
The production of acoustic signals from the interactions of ultra-high
energy (UHE) cosmic ray neutrinos in water and ice has been studied. A new
computationally fast and efficient method of deriving
the signal is presented. This method allows the implementation of
up to date parameterisations of acoustic attenuation in sea water and 
ice that now includes the effects of complex attenuation, where appropriate.
The
methods presented here have been used to compute and study the properties of
the acoustic signals which would be expected from such interactions. A
matrix method of parameterising the signals, which includes the expected
fluctuations, is also presented. These methods are used to generate the
expected signals that would be detected in acoustic UHE neutrino
telescopes.
\end{abstract}
\end{frontmatter}

\section{Introduction}
In recent years interest has grown in the detection of very high energy cosmic 
ray neutrinos which offer an unexplored window on the Universe \cite{ARENAs}. 
Such particles may be produced in the cosmic particle accelerators which make 
the charged primaries or they could be produced by the interactions of the 
primaries with the Cosmic Microwave Background, the so called GZK effect 
\cite{GZK}. 
The flux of neutrinos expected from these two sources has been calculated 
\cite{WBlim,GZKnu}. This is  found to be very low so that large targets are 
needed for a measurable detection rate. 
It is interesting to measure
this neutrino flux to see if it is compatible with the values expected from 
these sources, with any incompatibility implying new physics.

Searches for cosmic ray neutrinos are ongoing in AMANDA \cite{AMANDA}, IceCube 
\cite{IceCube}, ANTARES \cite{Antares}, NESTOR \cite{NESTOR}, 
NEMO \cite{NEMO}, KM3NeT \cite{KM3NET} and at Lake Baikal \cite{Baikal}
detecting 
upward going muons from the Cherenkov light in either ice or water. 
In general, these experiments are sensitive to lower energies than discussed 
here since the Earth becomes opaque to neutrinos at very high energies. The 
experiments could detect almost horizontal higher energy neutrinos but have 
limited target volume due to the attenuation of the light signal in the media. 
The Pierre Auger collaboration,using an extended air shower array detector, 
are searching for upward and almost horizontal showers from neutrino interactions 
\cite{Auger}. 
In addition to these detectors there are ongoing experiments to detect the 
neutrino interactions by either radio or acoustic emissions from the resulting 
particle showers \cite{ARENAs}. 
These latter techniques, with much longer attenuation lengths, allow very 
large target volumes utilising either large ice fields or dry salt domes for 
radio or ice fields and the oceans for the acoustic technique.

In order to test the feasibility of detecting such neutrinos by the acoustic 
technique it is necessary to understand the production, propagation and 
detection of the acoustic signal from the shower induced by an interacting 
neutrino in a medium. This has been treated in some detail in
\cite{Learned}, however, in this treatment it is difficult to incorporate the 
true attenuation of the sound 
which has been found to be complex in nature \cite{complex} 
in media such as sea water.
Such complex 
attenuation causes dispersion of the acoustic signal and complicates both the 
propagation of the sound through the water and the signal shape at the 
detectors.

This paper is organised as follows. In section 2 the new approach to calculating 
the acoustic signal pressure is described and section 3 describes the 
methods used to model the attenuation of the sound as it propagates through 
the medium. Section 4 describes the detailed calculations of the sound signal 
in water and in ice as it arrives at the detector starting from the shower
simulations described in \cite{ACORNE}. Finally a new method of simulating 
signals incorporating shower-to-shower fluctuations is described in Section 5.
\section{The thermo-acoustic mechanism} 
The standard equations used to determine the thermo-acoustic integrals 
are outlined in~\cite{Learned}.
In this paper we use a complementary approach. 
 
For the thermo-acoustic mechanism even though it is the pressure, 
$p$, that is detected it is the volume change, $Q_s$, which couples to the velocity potential, 
$\Phi$, which in turn creates the sound wave. 
\footnote{This is analogous to the situation in the
radio detection of UHE neutrinos where even though it is the electric field $E$ that is detected, 
it is the current density, $j$, which couples to the magnetic vector potential, $A$, 
which in turn creates the electromagnetic wave \cite{Olaf}.}
Interestingly, the velocity potential as a concept 
precedes the magnetic vector potential by over 100 years and was introduced by Euler in 1752~\cite{Euler}.

Three of the most important variables in acoustics 
are the pressure change from equilibrium, \textit{p}, the particle velocity, \textit{v}, and the 
velocity potential, $\Phi$. Assuming zero curl these three variables 
are related by:
\begin{equation}
v = \nabla \Phi, \quad \quad p = - \rho \frac{{\partial \Phi }}{{\partial t}},
\end{equation} 				
where $\rho$ is the density. For sources of acoustic energy this velocity potential has a function in 
acoustics equivalent to the magnetic vector potential in electromagnetism. 
We are trying to solced the wave equation to get the pressure pulse at the location of an observer placed at $|\overline{r}|$.
This can be dome by integrating all the contributions of infinitessimally small sources at locations  $|\overline{r'}|$.

\begin{figure}
\includegraphics[width=\textwidth]{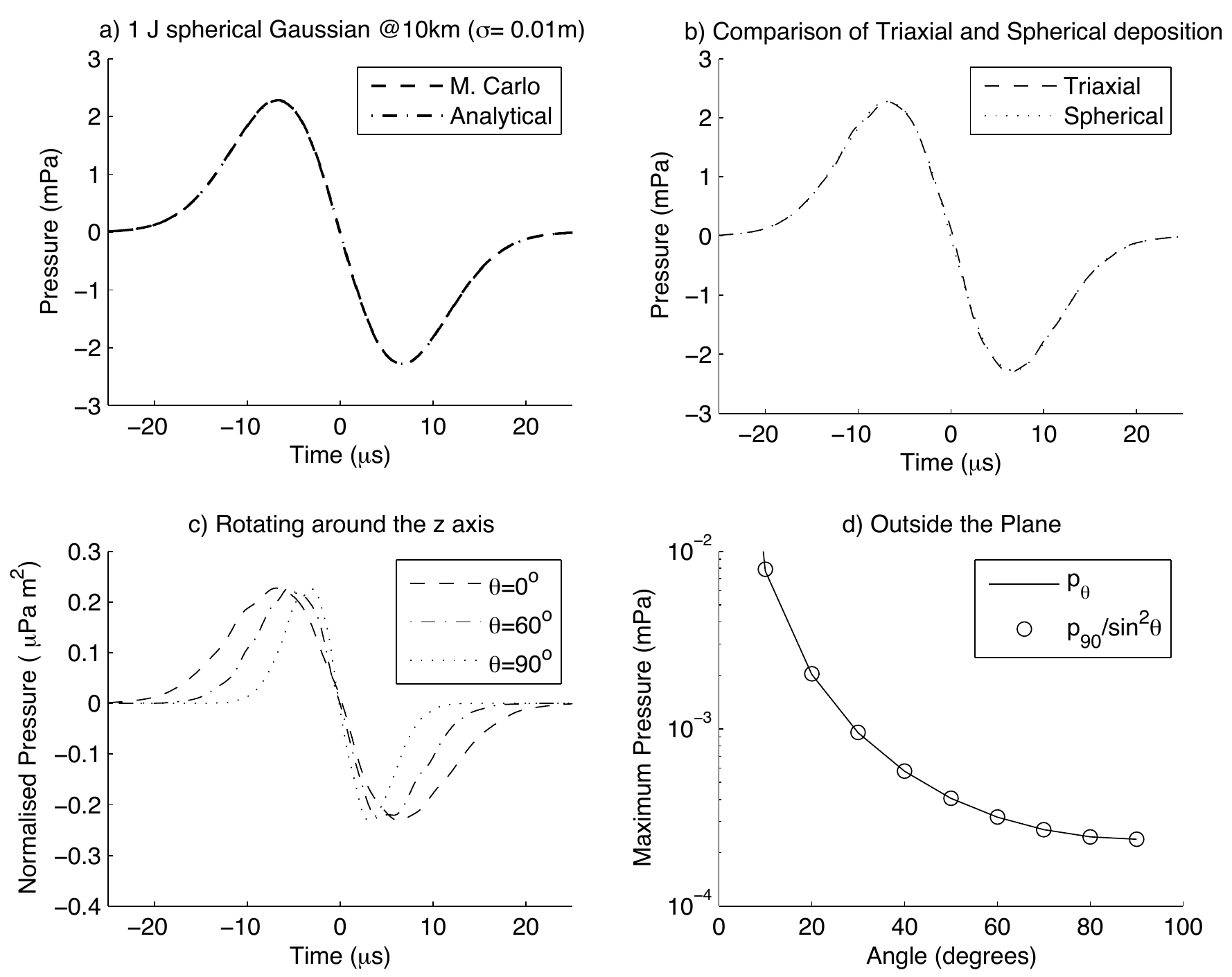}
\caption{Acoustic pulses from a 1~J deposition with an observer at 10 km,
a) comparison of analytical and MC pulse for a spherical deposition $\sigma$ =1 cm
b) comparison of a spherical ($\sigma$ =1 cm) and triaxial deposition ($\sigma_x$ =1 cm, $\sigma_y$ =0.5 cm, $\sigma_z$ =1 m)
c) rotating the observer from $x$-axis to the $y$-axis in the $x-y$ plane for the triaxial distribution
d) rotating the observer in the $x-z$ plane from the $x$-axis to the $z$-axis for the triaxial distribution.}
\label{fig1}
\end{figure}

For an observer at $\overline{r}$ and a shower event at $\overline{r'}$ separated by a distance $r = |\overline{r}-\overline{r'}|$:
\begin{equation}
\nabla ^2 \Phi - \frac{1}{{c^2 }}\ddot \Phi = \int q_s(r',t') dt',
\end{equation}
where $q_s$ is the time rate of change of an infinitesimal volume and $c$ the velocity of sound. In our case 
we are interested in integrating $q_s$ over the cascade volume where this volume change is caused by the injection of an energy density $E$ (in $J.m^{-3}.s^{-1}$) over this cascade volume. For an infinitesimal volume the volume change starting at time $t=0$, is given by:
\begin{equation}
q_s (\overline{r'},t) = \frac{\beta }{{\rho C_p }}E(\overline{r'},t) - \frac{1}{\tau }\int\limits_0^t {q_s (\overline{r'},t') dt'}, 
\end{equation}
where $\beta$ the thermal expansion coefficient, $C_p$ the specific heat capacity, $\rho$ the density 
and $\tau$ the thermal time constant. The integral term is caused by cooling as the deposited energy, 
within the volume, conducts or convects away into the surrounding fluid. However as the time constant 
for this thermal cooling mechanism is of the order of tens of milliseconds \cite{Underwood04}, and as 
we are primarily interested in the case where the energy is injected nearly instantaneously in 
acoustic terms (ns), the second term in equation 3 can be ignored as it is about six orders of magnitude lower than the 
first term. Equation 3 can then be integrated over the entire volume cascade using Green's functions (see for example \cite{Pierce81}) and yields 
for an observer at distance, $r$, from the source:

%
\begin{equation}
\Phi (\overline{r},t) = \int \int_{V}\frac{1}{|\overline{r}-\overline{r'}|}E(\overline{r'},t')_{(t=t'+|\overline{r}-\overline{r'}|/c)}dVdt'
\end{equation}
where $t-r/c$ is the retarded time, this simply implies that the contribution from each point in the source, 
in both space and time, travels to the listener with the speed of sound. Equation 4 lends itself to 
efficient numerical solution. If the energy deposition is modelled using Monte Carlo points with a 
density proportional to energy, $\Phi(t)$ will be a scaled histogram of the flight times to 
the observer. The pressure can now be derived from equation 1. A further simplification can be 
made if the energy deposition as a function of time is identical at all points in the volume. The 
velocity potential can then be calculated from the convolution integral:
\begin{equation}
\Phi (t) = \frac{\beta }{{4\pi C_p \rho }}\int {\int\limits_V {\frac{1}{r}E_{s}(\overline{r'})dV\;E_{t}(t^{'} - \lambda )d\lambda } }, 
\end{equation} 
where $\lambda$ is a dummy variable used to evaluate the integral. 
This can be simplified further since $E(t)$ can be approximated 
by $\delta(t)$ for the case we are interested in. The time dependence on the right hand side will disappear as the integral of a 
delta function is equal to one. This formulation of the thermo-acoustic mechanism leads
to a solution in the far field which is proportional to the Gr\"uneisen coefficient $\gamma_G$~\cite{Gruneisen}. If, after an arbitrary 3D 
rotation, the observer is, for example, at a distance from the source along the $x'$ axis much further 
away than the dimensions of the source, then with increasing observer distance, $x'_0$, the $E(y')$ 
and $E(z')$ contribution will more and more closely resemble delta functions as the spread in 
arrival times caused by their contributions approaches zero. Equation 5 reduces to:
\begin{equation}
\Phi (t) = \frac{\beta }{{4\pi C_p \rho }}\frac{c}{{x'_0 - x'}}E(c(t-t_{0})) \approx \frac{\beta }{{4\pi C_p \rho }}\frac{c}{{x'_0 }}E(c(t-t_{0})),
\end{equation}
where $t_0$ is the flight time from the centre of the shower to the observer. The velocity potential, 
$\Phi(t)$, is simply a scaled cross section of the energy deposition and $p(t)$ a projection
of $\Phi(t)$:
\begin{equation}
\begin{array}{lcr}
p(t - t_0 ) = - \frac{\beta }{{4\pi C_p }}\frac{c}{{x'_0 - x'}}\frac{d}{{dt}}E(c(t-t_{0})) \\ = \frac{\beta }{{4\pi C_p }}\frac{{c^2 }}{{x'_0 - x}}\frac{d}{{dx'}}E(x') \\ \approx \frac{\beta }{{4\pi C_p }}\frac{{c^2 }}{{x_0 }}\frac{d}{{dx'}}E(x').
\end{array}
\end{equation}
The $\beta c^2/C_p$ term is the Gr\"uneisen coefficient and gives the relative acoustic pulse heights for different media. 
With this formulation the far field solution in the absence of attenuation makes a number of predictions:
\begin{enumerate}
\item[a)]{The pressure pulse is a scaled derivative of the projection along the line of sight to the observer;} 	
\item[b)]{Only the distribution along the line of sight is important. Hence, for example, with an observer on the 
$x$ axis, a 1 J deposition into a Gaussian sphere (i.e. where $x$,$y$ and $z$ are randomly distributed with a normal distribution
with the same standard deviation, $\sigma_{x,y,z}$)
with $\sigma=1$ cm, centred on the origin, 
will observe the same pulse as a tri-axial Gaussian distribution 
(as above where the standard deviations in $x,y$ and $z$ are different)  
with $\sigma_x=1$ cm, $\sigma_y=1$ mm and 
$\sigma_z=1$ m centred on the origin. This is the case when $x >> \sigma_y,\sigma_z$;}
\item[c)]{The amplitude of the pressure pulse will depend on $1/\sigma_0^2$ where $\sigma_0$ is the projected 
standard deviation in the acoustic pulse height as seen from the observer. Hence an observer on the $y$ axis will see a pulse one 
hundred times greater in magnitude than one on the $x$ axis for the tri-axial distribution as
described in item b) above;}
\item[d)]{At angles greater than a few degrees from the $x-y$ plane the amplitude of the pressure pulse will depend 
on $1/\sin^2\theta$.}
\end{enumerate}
This is illustrated in figure~\ref{fig1} where each of the sub-figures illustrate
the corresponding point in the list above.
\section{Attenuation}
\subsection{Introduction}
The acoustic pulse is affected by the medium through which it travels and
it acts as a filter causing frequency dependent attenuation, $a(\omega)$. Attenuation of acoustic pulses is 
caused by a combination of absorption and back scattering. The amplitude is attenuated by a factor given by:
\begin{equation} 
a(\omega ) = e^{ - k(\omega )r},
\end{equation}
\noindent 
where $r$ is the distance travelled and $k$ is the attenuation coefficient (Nepers/m). The resultant pulse can be determined by converting the 
pulse into the frequency domain by taking a Fourier transform, multiplying and taking the inverse Fourier 
transform:
\begin{equation}
\Phi _a (t') = \frac{1}{{2\pi }}\int\limits_{ - \infty }^\infty {a(\omega )} \left( {\int\limits_{ - \infty }^\infty {\Phi (t')e^{ - i\omega t} dt} } \right)e^{i\omega t} d\omega,
\end{equation}
where $\Phi_a$ is the velocity potential with attenuation and $t'=t-r/c$. Alternatively the effect of 
attenuation can be determined by taking the inverse Fourier transform of the frequency 
characteristic (i.e. the modification of the pulse between transmission and reception due to the medium)
and convolving this with the un-attenuated pulse:
\begin{equation}
\begin{array}{lcr}
 a(t) = \frac{1}{{2\pi }}\int\limits_{ - \infty }^\infty {a(\omega )} e^{i\omega t} d\omega \\ 
 \Phi _a (t') = \int\limits_{ - \infty }^\infty {a(t' - \lambda )\Phi (\lambda )d\omega } = \int\limits_{ - \infty }^\infty {\Phi (t' - \lambda )a(\lambda )d\lambda }, \\ 
 \end{array}
 \end{equation}
\noindent 	
where $\lambda$ is a dummy variable used to evaluate the integral and $a(t)$ is the pulse created by passing a 
Dirac delta function source through the medium and is referred to as the unit impulse response 
The equivalence of these methods is a statement 
of the convolution theorem: multiplication in the frequency domain is convolution in the time domain.
Values of the pressure attenuation length in sea water (ice) are 3.8~km (10~km) at 10~kHz and 2.6~km (8.5~km) at
25~kHz. Attenuation lengths for intensity are a factor of two less.
\subsection{Attenuation in distilled water}
Evaluating $a(t)$ analytically is normally difficult. However for a liquid in which the attenuation is 
dominated by viscosity it is straight forward. This is the case for example in distilled water. It can 
be shown (see, e.g.~\cite{Learned}) that in this case the attenuation coefficient, $k$, and 
attenuation, $a$, for frequencies of interest, are given by:
\begin{equation} 
k = \frac{1}{{2c}}\frac{{\omega ^2 }}{{\omega _0 }},\quad a(\omega ) = e^{ - \frac{1}{{2c}}\frac{{\omega ^2 }}{{\omega _0 }}r} = e^{ - \frac{{\omega ^2 }}{{2\sigma _a(\omega) ^2 }}} ~\rm{where}~ \sigma _a(\omega) = \sqrt {\frac{{c\omega _0 }}{r}},
\end{equation}
and  $\omega_0$ is medium dependent. 
The attenuation $a(\omega)$ is Gaussian in shape with a standard deviation of $\sigma_a$ which is proportional to $1/\sqrt{r}$. 
In distilled water, $\omega_0 \approx 10^{12}$ 
rad s$^{-1}$ and Lehtinen et al.~\cite{SAUND} have used a value of $2.5 \times 10^{10}$ rad s$^{-1}$ 
to approximate the attenuation of tropical sea water in the $10^4-10^5$ Hz region. 
The unit impulse response is given by the inverse Fourier Transform as:
\begin{equation} \begin{array}{lcr}
 a(t) = \frac{1}{{2\pi }}\int\limits_{ - \infty }^\infty {e^{ - \frac{r}{{2c\omega _o }}\omega ^2 } } e^{i\omega t} d\omega = \sqrt {\frac{{\left( {\frac{{c\omega _o }}{r}} \right)}}{{2\pi }}} e^{ - \frac{{c\omega _o }}{{2r}}t^2 } \\ 
 = \frac{1}{{\sqrt {2\pi } \sigma _a }}e^{ - \frac{{t^2 }}{{2\sigma _a(t)^2 }}}  ~\rm{where}~ \sigma _a(t) = \sqrt {\frac{r}{{\omega_o c}}}, \\ 
 \end{array} \end{equation}
\noindent 
and is also a Gaussian distribution with a standard deviation which gets wider with the square root 
of the distance. Indeed $\sigma_a(t)=1/\sigma_a(\omega)$.
If the source term is also Gaussian 
the overall pulse profile can be evaluated. The convolution integral of two Gaussian distributions 
yields a Gaussian distribution with the standard deviations adding in quadrature. Therefore if:
\begin{equation} 
\begin{array}{lcr}
 \Phi (t') = \frac{{\beta E}}{{4\pi C_p \rho r}}\frac{1}{{\sqrt {2\pi } \sigma _\Phi }}e^{ - \frac{{t'^2 }}{{2\sigma _\Phi ^2 }}} \\ 
 \Phi _a (t') = \frac{{\beta E}}{{4\pi C_p \rho r}}\frac{1}{{\sqrt {2\pi } \sigma _{\Phi _a } }}e^{ - \frac{{t'^2 }}{{2\sigma _{\Phi _a }^2 }}} ~\rm{where}~ \sigma _{\Phi _a } = \sqrt {\sigma _{^{_\Phi } }^2 + \sigma _{^{_a } }^2 }, \\ 
\end{array} 
\end{equation} 
then the pressure $p$ is given by:
\begin{equation} 
p(t) = \rho \frac{d}{{dt}}\Phi _a (t') = - \frac{{\beta E}}{{4\pi C_p r}}\frac{{t'}}{{\sqrt {2\pi } \sigma _{_{\Phi _a } }^3 }}e^{ - \frac{{t'^2 }}{{2\sigma _{\Phi _a }^2 }}}. 
\end{equation}
This is equivalent to equation 18 in reference \cite{Learned}, but is derived using a quite different approach.
\subsection{Attenuation in sea water and ice}
%
Acoustic attenuation in seawater is almost totally caused by
absorption. In the 1-100kHz region it is dominated by a chemical
relaxation process that is connected with the
association-dissociation of magnesium sulphate ($MgSO_4$) ions under the
pressure of the sound wave. Below 1kHz a similar mechanism involving
boric acid ($B(OH)_3$) is responsible for much of the observed
attenuation. Taken together these mechanisms result in an attenuation of acoustic waves
in seawater and a velocity of sound which are both frequency dependent.
Experimentally however whereas it is straightforward to estimate the magnitude of $|z|$ 
of the complex attenuation it is very difficult to 
determine the phase angle $\angle z$ and no current measurements exist in the literature. 

The magnitude 
of the attenuation however is well measured and the definitive work in this area is 
considered to be that of Francois and Garrison \cite{FrancoisGarrison}. 
More recently, Ainslie and McColm \cite{AinslieMcColm} have published 
a simplified parameterisation of the magnitude of the attenuation in sea water which maintains
a similar accuracy to the parameterisations of Francois and Garrison. 
This is a function 
not only of frequency but also depends on depth, $z$, salinity, $S$, temperature, $T$ and 
pH. Whereas there are no direct measurements of the attenuation angle, 
Lieberman \cite{Lieberman} gives a clear presentation of
the chemical processes causing the attenuation while Niess and Bertin \cite{NiessBertin} have published 
a complex attenuation formula based on Mediterranean conditions. Here we present a complex version 
of the Ainslie and McColm formulation, which retains the attenuation magnitude but introduces the 
phase shifts predicted by Lieberman. In essence the attenuation consists of three components, two 
of these are complex, high pass filters with cut off frequencies $\omega_B$ ($\sim$2$\pi\times10^{3}$  rad s$^{-1}$) for boric 
acid and $\omega_{Mg}$ ($\sim$2$\pi\times10^{5}$ rad s$^{-1}$) for magnesium sulphate. The third is the pure water component, which is real. The
ACORNE parameterisation uses $a_x$ values that are the respective attenuation coefficients in dB/km:
\begin{equation}
\begin{array}{lcr}
 \omega _{B} = 1560\pi \sqrt {\frac{S}{{35}}} {\kern 1pt} e^{T/26} \quad, \quad \omega _{Mg} = 84000\pi {\mathop{\rm e}\nolimits} ^{T/17} \\ \\
 a_B = \frac{{{\rm{1}}{\rm{.893}} \times {\rm{10}}^{{\rm{ - 4}}} }}{{{\rm{2}}\pi }}{\mathop{\rm e}\nolimits} ^{\frac{{pH - 8}}{{0.56}}} {\rm{ ,}}\quad a_{Mg} = \frac{{0.52 \times {\rm{10}}^{{\rm{ - 3}}} }}{{{\rm{2}}\pi }}\left( {1 + \frac{T}{{43}}} \right)\frac{S}{{35}}e^{\frac{{ - z}}{6}} \\ \\ 
 a_w = \frac{{49 \times {\rm{10}}^{{\rm{ - 9}}} }}{{4\pi ^2 }}{\mathop{\rm e}\nolimits} ^{ - (T/27 + z/17)} \\ \\ 
 a_{dB/km} = \frac{{a_B \omega _B s}}{{s + \omega _B }} + \frac{{a_{Mg} \omega _{Mg} s}}{{s + \omega _{Mg} }} + a_w \omega ^2 \quad ~\rm{where}~ {\kern 1pt} s = i\omega. 
\end{array}
\end{equation}
%
In ice the mechanisms are less well understood. However following Price~\cite{Price} in regions of deep ice 
of most interest for acoustic neutrino detection, e.g. the South Pole, the dominant attenuation mechanisms 
below a few hundred kHz are: absorption due to proton reorientation (relaxation) and scattering due to 
bubbles and grain boundaries. For South Pole conditions Price predicts the absorption length 
(apart from very low frequencies $<$100Hz) to be constant:
\begin{equation}
a_L (m^{-1}) = \frac{\delta_{max}}{\nu_L}\frac{\omega^2\tau_m}{\left({1+\omega^2\tau_m^2}\right)},
\end{equation}
where $\delta_{max}$ is the log decrement at asymptotically high frequencies, $\tau_m$ is a temperature 
dependent relaxation time and $\nu_L$ is the velocity of sound in ice (3920 m s$^{-1}$).
The parameters are normally evaluated experimentally. For South Pole conditions these predict an absorption length of 
$\sim$10 $\pm$3 km at frequencies above 100 Hz. 
The dominant scattering mechanism is caused by grain boundaries and for the grain size expected at the South Pole the ice
will behave as a Rayleigh medium. Scattering is proportional to the 4th power of frequency,
up to $\sim$300 kHz:
\begin{equation} 
b_L (km^{-1}) = 5 \times 10^4 \left( {\frac{d}{{0.2{\rm{cm}}}}} \right)^3 \left( {\frac{f(kHz)}{{10{\rm{kHz}}}}} \right)^4,
\end{equation}
where $d$ is the mean grain diameter which is 0.2~cm. As the acoustic pulses from neutrino interactions tend to be 
highly directional we make the assumption that the attenuation is given by the sum of the absorption 
plus total scattering rather than the more usual assumption that the attenuation is given by the 
sum of absorption plus back-scattering.

\begin{figure}
\includegraphics[width=\textwidth]{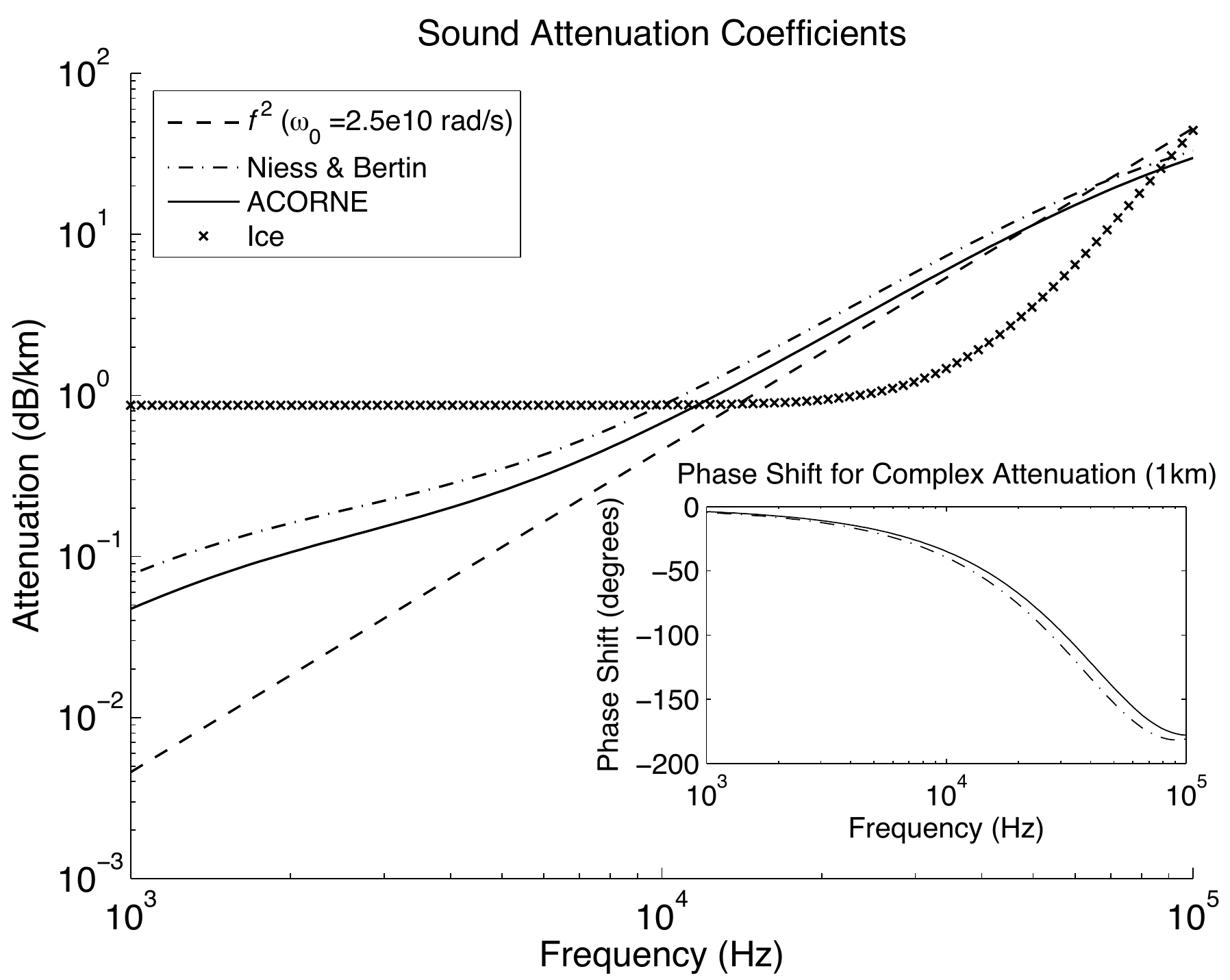}
\caption{Sound attenuations in sea water for the SAUND/Learned ($f^2$, Niess and Bertin, Francois and Garrison, 
Ainslie and McColm and the ACORNE parameterisations. Since the Ainslie \& McColm/ACORNE and Francois \& Garrison 
results match very closely they are depicted as the same curve in the figure.
The attenuation in Antarctic ice is shown for comparison. 
Inset: the phase shifts at 1 km for the two complex attenuation models.}
\label{fig2}
\end{figure}

A comparison of the ACORNE attenuation parameterisation and earlier work is presented in figure~\ref{fig2}, 
together with the anticipated attenuation in ice.  The 
$f^2$ approximation assumes the sea behaves like a more viscous form of distilled water by decreasing 
the value of $\omega_0$ to match the measured attenuation in the $10^4$ to $10^5$ Hz frequency region for tropical 
waters, consistent with the location of the SAUND array \cite{SAUND}. The other curves are optimised for Mediterranean 
conditions ($T = 15^{\circ}\rm{C}; S = 37; \rm{pH} = 7.9; z = 2 \rm{km}$). 
Since the Ainslie \& McColm/ACORNE and Francois \& Garrison results match very closely they are depicted as the same 
curve in the figure. 
Phase shifts for a pulse 1 km distant from the neutrino interaction
are shown for the complex attenuation case, corresponding 
to a velocity increase of $\sim$0.005\% between 1 and 100 kHz. For ice the attenuation is constant up 
to a frequency $\sim$25 kHz where Rayleigh scattering starts to dominate. The absorption in ice is lower than that of 
sea water in the important 10-100 kHz region. 

\begin{figure}
\includegraphics[width=\textwidth]{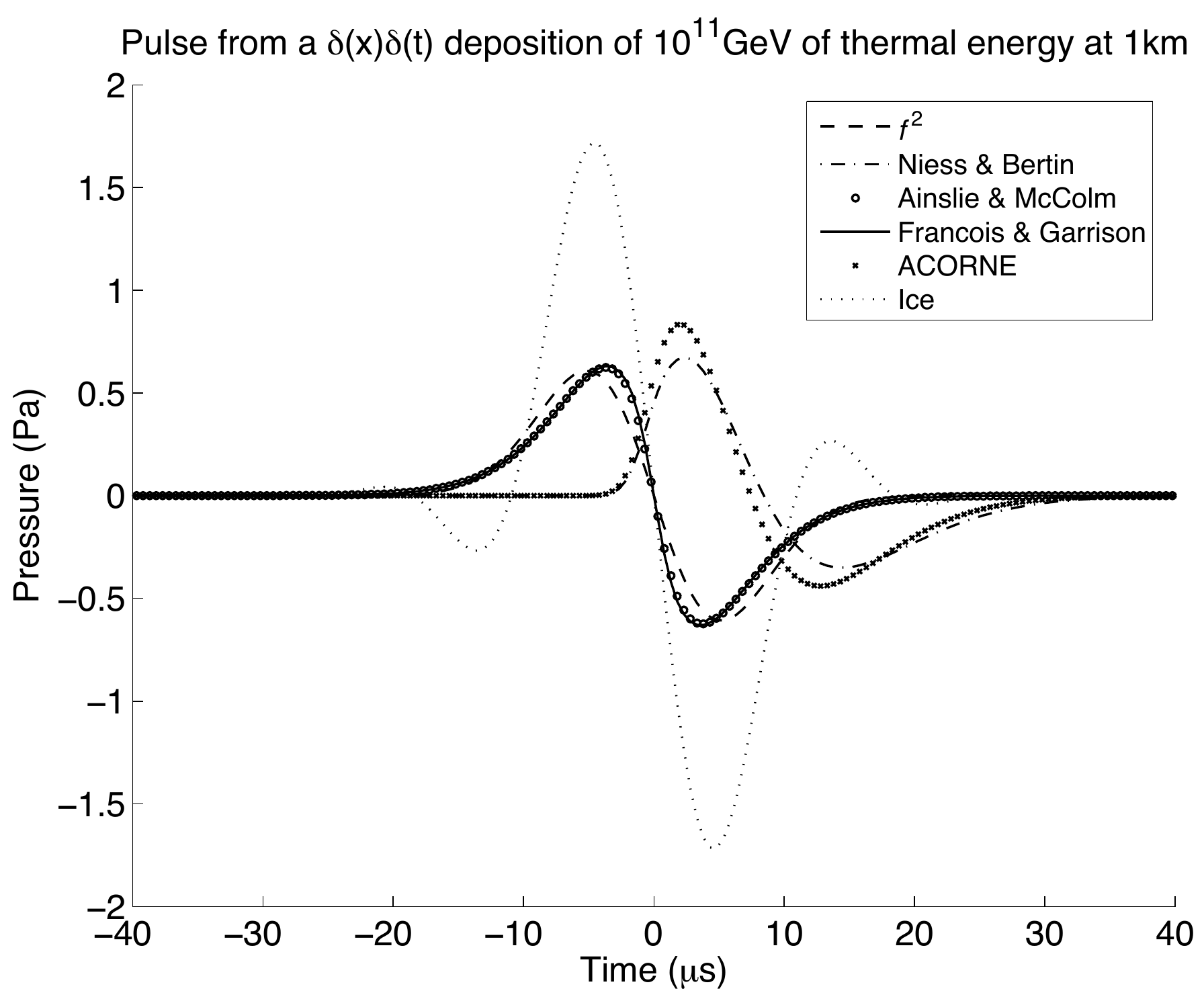}
\caption{The acoustic pulse for a point deposition of energy and an observer at 1km from the source for the  $f^2$, Niess and Bertin, 
Francois and Garrison, Ainslie and McColm and ACORNE parameterisations. 
The anticipated acoustic pulse for Antarctic ice is shown for comparison. The ACORNE and Niess and Bertin pulses are delayed due to the use
of complex attenuation.}
\label{fig3}
\end{figure}

The effect of the various forms of attenuation on a point source of energy $\delta (r) \delta (t)$ 
observed at 1 km from the origin is shown in figure~\ref{fig3}. The three real attenuation mechanisms in water 
give similar symmetric pulses. This gives confidence that the magnitude of the attenuation need not 
be modelled in fine detail. Interestingly however whereas the magnitude of the two complex attenuation 
models is similar to the other water models, (and, indeed identical in the case of the ACORNE and 
Ainslie \& McColm models) the phase shifts cause considerable pulse distortion from the anticipated 
symmetric shape. This is a result of the non-linear nature of the phase shifts creating the group delay, 
$d\phi /d\omega$, to vary, hence creating dispersion. The phase shift also has the effect of marginally 
reducing the pulse velocity at lower frequencies hence slightly delaying the pulse. 
In ice the pulse is again symmetric and is larger in amplitude by a factor of three. 
This is not, as one might expect, due to the larger value of the Gr\"uneisen coefficient
in ice, but is a direct consequence of the
lower attenuation in the 10-100 kHz region. \footnote{The ratio of
$\beta/C_p$ is 1.38 between ice and water and the $1/c^2$ term is a factor of 6.83 larger for ice.
However, the $1/c^2$ term 
only applies to extended sources and can be ignored for a point source.}
The ripples before and after the 
main pulse are caused by the abrupt $f^4$ scattering dependence switching on yielding a rapid rise in 
attenuation. Any sharp transition in the frequency domain will tend to cause ripples in the time domain,
these ripples are a real physical effect and not a numerical artefact.
\section{ Acoustic integrals for CORSIKA simulated pulses in water and ice}
\subsection{Introduction}
In this section we compute full 3D simulations of acoustic pulses from our 
previously published neutrino shower simulations \cite{ACORNE}. In our previous work, 1500 showers were modelled using the air shower
simulation program CORSIKA \cite{CORSIKA} modified for a water or ice medium \cite{ACORNE} corresponding to 100 showers in each 
half decade increment from $10^5$ to $10^{12}$ GeV. 
The deposited energy was binned in a cylinder of 20 m in length and 1.1 m in radius. 
Longitudinally 100 bins of width 20 cm with bin centres from 10 cm to 1990 cm were used. Radially, 20 bins were used: ten of 
width 1 cm with centres from 0.5 to 9.5 cm and 10 of width 10 cm  with centres from 15 cm to 105 cm. The contribution of the outer 
ten bins towards the acoustic pulse is minimal; they are included to capture all the energy in the shower and are useful for conservation 
of energy checks. This spacing should be sufficient to accurately model the acoustic frequency components in the pulse on axis 
up to $\sim$75 kHz in water and $\sim$200 kHz in ice. 
In this section the average shower distribution for each of the 15 half decades in energy is used directly as an input to drive the acoustic integral. 
The neutrino creates a hadronic shower of approximately 10 m in length and 5 cm in radius. 
The thermal energy is effectively deposited instantaneously in acoustic terms, creating 
an acoustic radiator analogous to a broad slit. In the far field, using Fraunhofer diffraction, the angular spread of the radiation is 
approximately $\lambda/d$. The wavelength $\lambda$ is of the order of the diameter of the cylinder and $d$ the length of the 
cylinder, yielding a pancake width in the order of 1 degree.
Water and ice are different media, the suitability of the medium for acoustic detection
will depend largely on the magnitude of the created acoustic pulse and the 
attenuation. 
For extended sources of this nature a suitable figure of merit to compare different media is given by the ratio of the product of the Gr\"uneisen coefficient ($\gamma_G$)
and the anticipated energy deposition ($dE/dX$):
\begin{equation}
\varepsilon  = \gamma_G \rho \frac{{dE}}{{dx}} = \frac{{c^2 \beta }}{{C_P }}\rho \frac{{dE}}{{dx}}.
\end{equation}
%
Using this expression, ice should produce acoustic pulses which are about an order of 
magnitude greater in amplitude than water.
However because the velocity of sound in ice is about 2.5 times that of water, the frequencies in the pulse will also be about 2.5 times higher 
causing ice to have a greater attenuation than water. 
%
\subsection{Coordinate system}
In the subsequent analysis a coordinate system is chosen such that the neutrino interacts at the origin and travels vertically along the $z$ axis where 
the value of $z$ increases with depth and the origin. The point $(0,0,-z_p)$ is chosen such that the maximum ``pancake'' energy 
(i.e. acoustic, and not deposited energy) at 1 km from the shower is at zero degrees. 
The value of $z_p$ is energy dependent varying from approximately 460 cm at $10^5$ GeV to 780 cm at $10^{12}$ GeV. 
If the radial cross-section of the shower were constant one would expect the pancake maximum to correspond to the centroid of the energy 
deposition taken along the longitudinal axis. In reality however the radial distribution gets broader with the age (depth) of the shower; 
the earlier part contributes more to the pulse energy than the later part of the shower. The point (0,0,-$z_p$) can be determined in two ways, 
either from the position of the shower centroid, $z_c$ or from $E$, the primary energy (in GeV) which defines the pancake plane. 
A fit to sea water data yields following relationships for $z_p$:
\begin{equation}
\begin{array}{l}
 z_p  = 1.05z_c  - 0.874 \\ 
 z_p  =  - 0.046\left( {\log _{10} E} \right)^2  + 1.3\log _{10} E - 0.84. \\ 
 \end{array}
\end{equation}
In Antarctic ice the pancake depth has to be increased by approximately 6\% due to the relative densities of the media.
\subsection{Acoustic integrals in sea water}
\begin{figure}
\includegraphics[width=\textwidth]{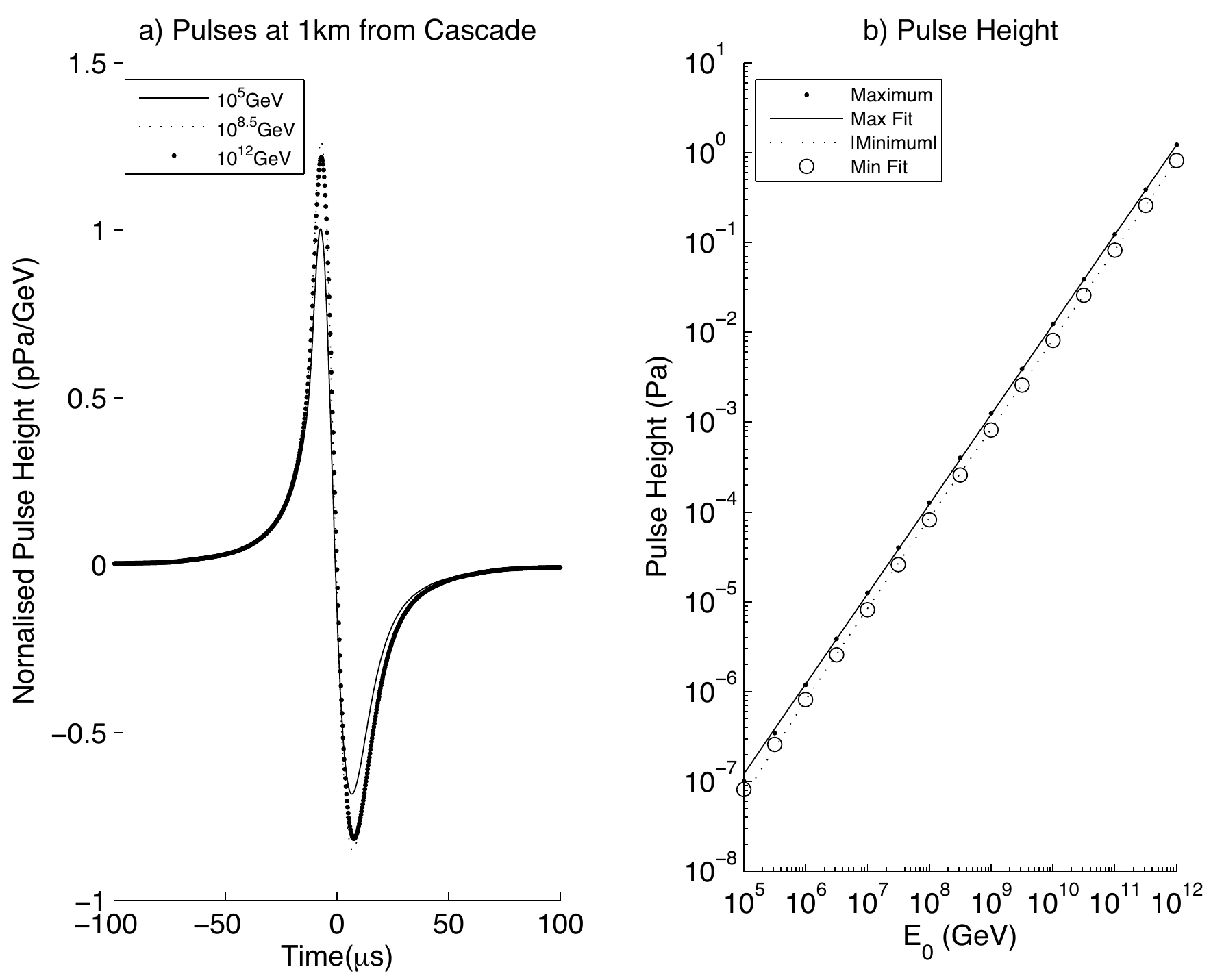}
\caption{The pulse at 1km for $10^{5}-10^{12} $GeV showers in water. 
a) pulse shape
b) maximum and minimum pulse heights vs. energy }
\label{fig4}
\end{figure}

For the initial analysis the observer is positioned at 1 km from the shower in the centre of the acoustic pancake to allow 
easy comparison with 
previously published results \cite{SAUND}\cite{NiessBertin}. Complex attenuation (equations 15) was assumed and the acoustic integral calculated for each of the 15 half decades 
in energy as described above. The results are plotted in figure~\ref{fig4}. As the integrals have been 
calculated with high precision, error-bars are not plotted, as they are less than a line width.
As can be seen a characteristic bipolar pulse is produced. In figure~\ref{fig4}a) the pulse shape is plotted for three energies. The maximum pulse amplitude 
is normalised by energy. As can be seen the pulse shape is very similar for the three chosen energies ($10^5$, $10^{8.5}$ and $10^{12}$ GeV). 
The pulse height however seems to scale slightly more rapidly than energy. In figure~\ref{fig4}b) the maximum and minimum pulse amplitudes are plotted 
on a log-log plot. The fitted lines are constrained to be proportional to energy. It is clear from figure~\ref{fig4}b) that any increase in pressure 
over proportionality is minimal. A linear fit yields:
\begin{equation}
\log _{10} (p_{\max } ) = 1.0021(\pm 0.0018)\log _{10} (E) - 11.93(\pm 0.0012),
\end{equation}
where $p_{max}$ (Pa) is the maximum pressure and $E$ the energy in GeV. The errors quoted are the statistical errors from the fit. 
This yields the result that, to a good approximation, 
the maximum pulse height at 1 km is 1.22 pPa per GeV in the plane of the pancake.
The average frequency (using the MUSIC algorithm \cite{MUSIC}) is very stable with energy and is approximately 26 kHz. 
The asymmetry $(|P_{max}|-|P_{min}|)$/$(|P_{max}|+|P_{min}|)$ is similarly stable and is about 0.2.

\begin{figure}
\includegraphics[width=\textwidth]{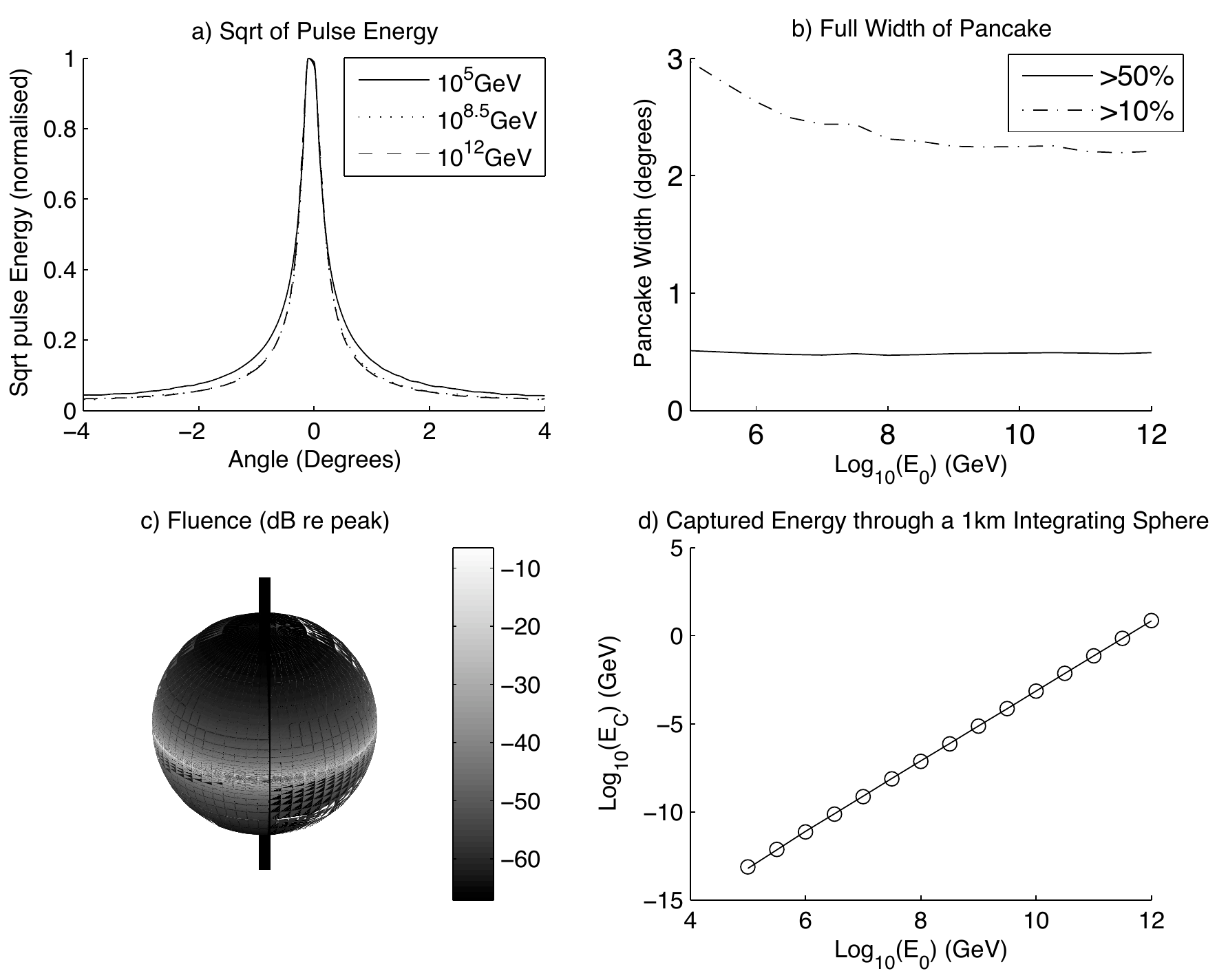}
\caption{Variation of acoustic pulse energy with angle in sea water
a) square root of the pulse energy versus angle
b) full width for 10\% and 50\% of maximum versus energy
c) fluence through an integrating sphere
d) captured energy versus deposited energy and linear fit (see equation 22).}
\label{fig5}
\end{figure}

Consider now figure~\ref{fig5} showing the angular spread of the pancake. As the acoustic pulse will most likely be detected by a matched filter 
(see, for example, ~\cite{SAUND} for a full discussion of matched filters), which integrates 
over the pulse, it is the integrated pressure, or square root of the pulse energy which is of most interest. This is plotted in figure~\ref{fig5}a), which 
shows the angular spread of the shower from $-4^\circ$ to $4^\circ$. The pancake is extremely narrow and narrows further with increasing energy. In figure~\ref{fig5}b) 
the full width of the pancake at a pressure levels of 50\% and 10\% of maximum is illustrated.  
As anticipated from simple far field diffraction theory (the diffraction minimum is at $\theta \tilde{=} \lambda/d$) 
the spread of the pancake decreases with increasing shower energy. This is largely because the shower gets longer as the energy increases causing a narrowing of the pancake. 

It is interesting to cross check the acoustic mechanism by looking at the energy flowing through a 1 km integrating sphere. The energy of the acoustic pulse 
flowing through each square metre of surface (the fluence) is given by:
\begin{equation}
\Phi (r,\theta ) = \int_{ - \infty }^\infty  {\frac{{p^2 (t,r,\theta )}}{{Z_0 }}dt},
\end{equation}
                      
where $Z_0$ is the characteristic impedance ($\sim$$1.5\times10^6$ kg m$^{-2}$ s$^{-1}$ for water and $\sim$$3.7 \times10^6$ kg m$^{-2}$ s$^{-1}$ for ice) and $\theta$ 
is the angle out of the plane of the pancake, the radiation is cylindrically symmetrical. This can be integrated over a sphere, 
in figure~\ref{fig5}c) the result of this integral is displayed. 
The fit of captured acoustic energy, $E_c$ vs. deposited energy $E_0$ (see figure~\ref{fig5}d))is given by:
\begin{equation}
\log _{10} E_c  = 2.00\log _{10} E_0  - k,
\end{equation}
where k= 22.8 without attenuation and 23.1 with attenuation. Hence at 1 km attenuation reduces the acoustic energy to 50\% of its value without attenuation. 
The linear coefficient is 2.00 to within the statistical accuracy of the showers generated by CORSIKA, indicating that the efficiency rises quadratically with energy. 
Due to the coherent nature of the acoustic emission mechanism,
the acoustic pulse amplitude depends linearly and the acoustic energy depends 
on the square of the deposited energy (assuming constant shower shape). 
This coherent behaviour breaks down at energies far beyond those of interest here.

\begin{figure}
\includegraphics[width=\textwidth]{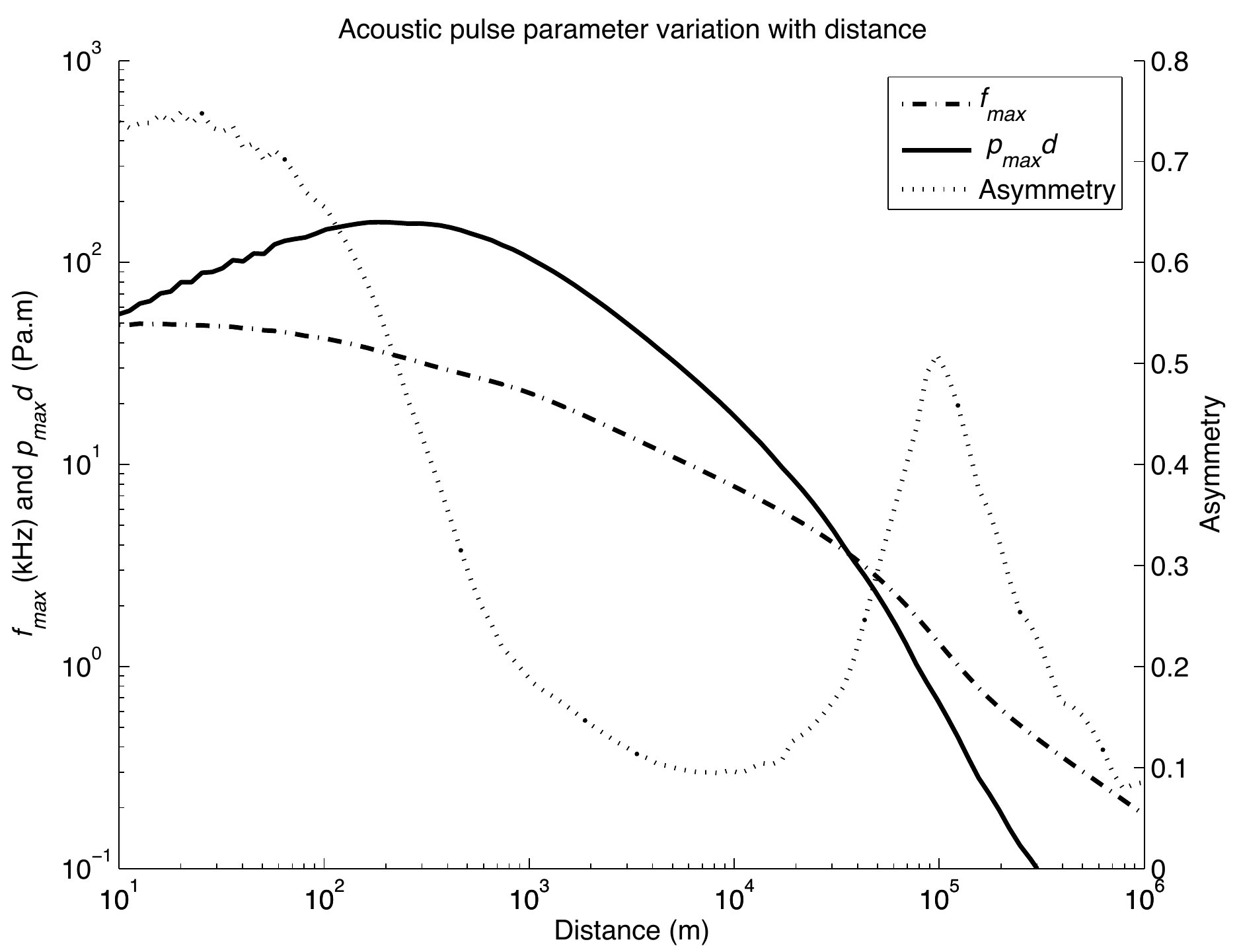}
\caption{Mean frequency, Pressure$\times$distance and asymmetry for the acoustic pulse in the pancake plane for a $10^{11}$GeV shower in sea water
plotted as a function of distance.}
\label{fig6}
\end{figure}

It is interesting to look at the effect of complex attenuation on the pulse asymmetry in water as there is no guarantee that the pulse will become symmetrical 
even in very far field. Figure~\ref{fig6} illustrates the mean pulse frequency, pressure times distance and asymmetry for a 10$^{11}$GeV energy deposition and 
for an observer in the pancake plane for distances of 10 m to $10^{3}$ km. In the absence of attenuation, once far field has been established ($\sim$1 km) both $p_{max}d$ 
and the mean frequency, $f_{max}$ should be constant and the asymmetry zero. 
In practice the mean frequency drops with distance, as higher frequencies 
are more quickly absorbed than lower frequencies, falling from $\sim$50 kHz close to the shower to $\sim$200 Hz at $10^{3}$ km. 
The product of maximum pressure and distance rises 
initially as the radiator approaches far field conditions. The maximum pressure then drops both because the energy is absorbed and the pulse becomes more 
spread out in time. The asymmetry starts with a value of 0.7 dominated by near field effects then drops to $\sim$0.1 at 1-10 km and has a peak of $\sim$0.5 at 
100 km. The dispersion will be a maximum at frequencies around the resonant peaks of MgSO$_4$ and B(OH)$_3$. The B(OH)$_3$ peak is at $\sim$1.3 kHz causing the 
rise in asymmetry where the mean frequency matches this value.

\begin{figure}
\includegraphics[width=\textwidth]{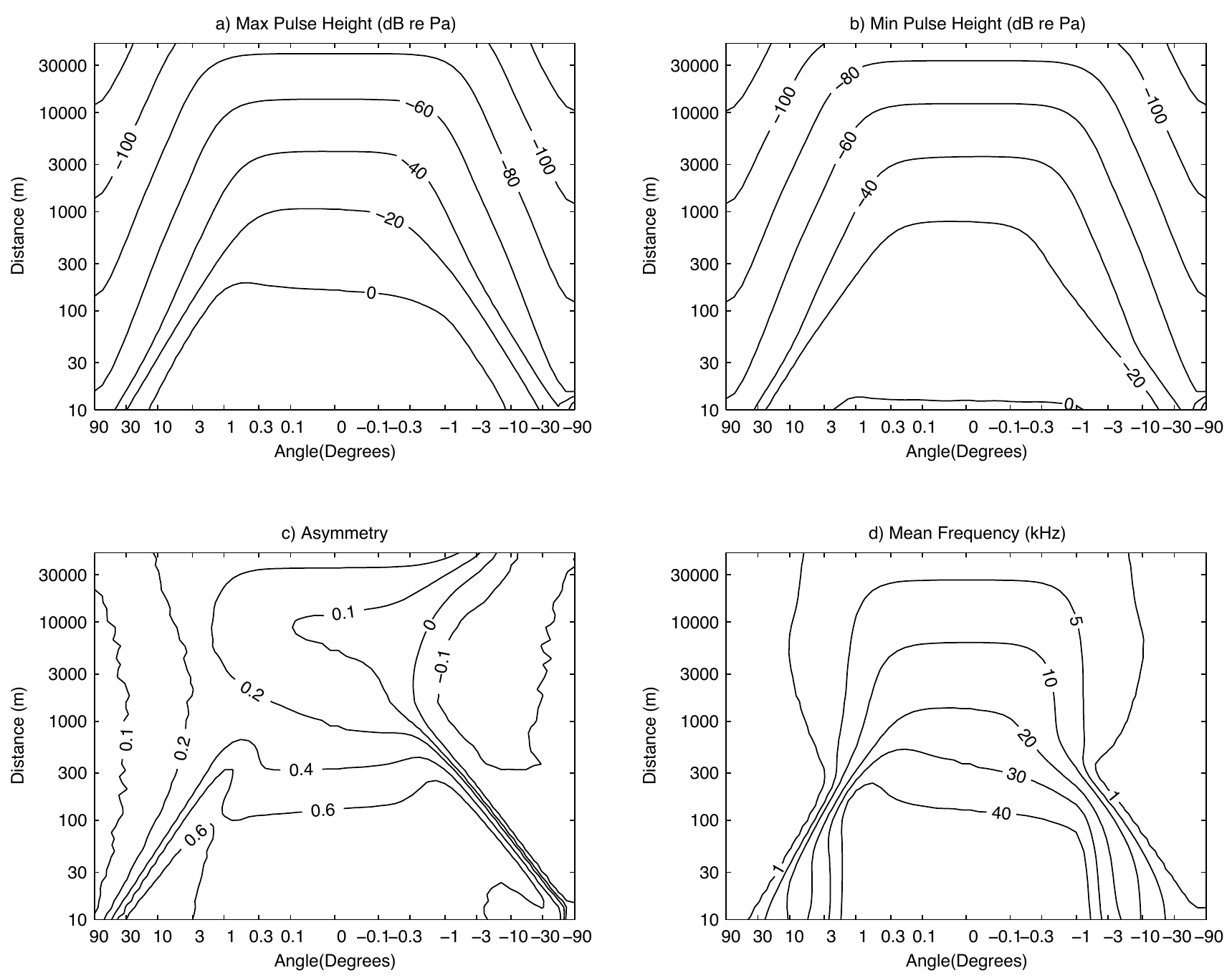}
\caption{Pulse parameters in sea water for a $10^{11}$GeV primary as a function of distance and angle
a) maximum pulse height
b) minimum pulse height
c) pulse asymmetry
d) mean frequency.}
\label{fig7}
\end{figure}

In figure~\ref{fig7} the mean frequency, 
maximum and minimum pressures and asymmetry are plotted versus angle and distance
for a 10$^{11}$ GeV shower using the ACORNE complex attenuation (T=15$^\circ$C, S=37ppt, pH=7.9, z=2km).
As the neutrino is travelling vertically downwards then positive angles (left hand side in figure~\ref{fig7}) are measured out of the plane of the
pancake and towards the surface of the water. 
In figures~\ref{fig7}a) and \ref{fig7}b) the maximum and minimum values of $P$ in dB re. 1 Pa are displayed. 
As anticipated, the pressure decreases with angle and frequency. 
Figure~\ref{fig7}c) shows the asymmetry where 
two effects are particularly noteworthy. Firstly the dominant 
effect of complex attenuation drives the pulse towards positive asymmetry. 
Secondly there are two regions with asymmetry of greater than 0.6. 
In these regions the geometry causes a spike in the velocity potential yielding a very non bipolar pulse. 
Consider now figure~\ref{fig7}d) showing the mean frequency. As can be seen at angles 
below $\sim$1$^\circ$ and distances below 100 m the frequency with maximum energy is above 40 kHz. This reduces with both angle and distance. The decrease of 
frequency with angle is caused mainly by geometric projection as discussed in section 2. The decrease of frequency with distance is caused mainly by 
high frequencies being attenuated more rapidly than low frequencies. 
%
\subsection{Acoustic integrals in ice}
\begin{figure}
\includegraphics[width=\textwidth]{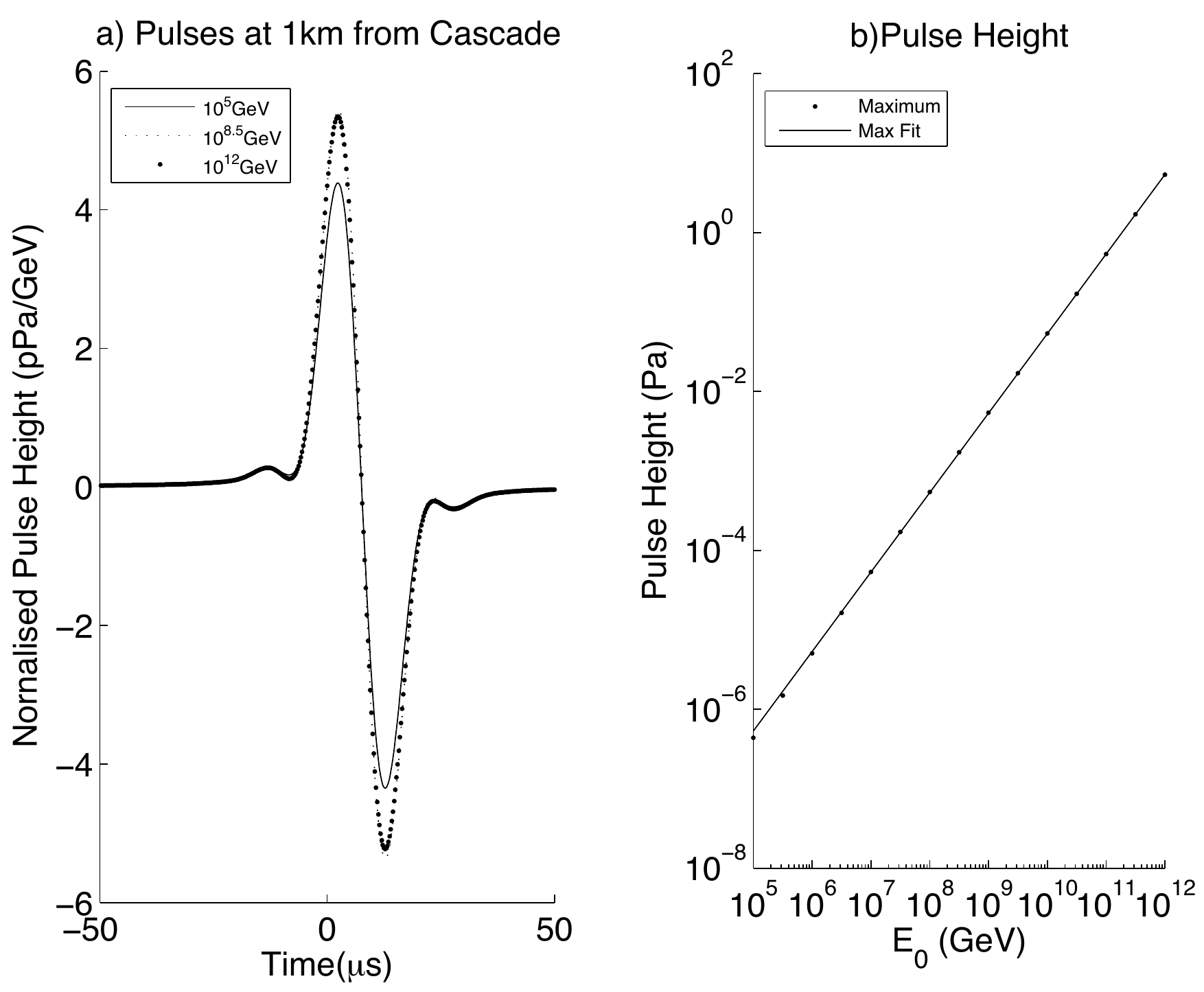}
\caption{The pulse at 1km for $10^{5}-10^{12} $GeV showers in ice. 
a) pulse shape, 
b) maximum and minimum pulse heights vs. energy.}
\label{fig8}
\end{figure}

Following the procedure outlined in section 3.2 and adopting a model of Antarctic ice from \cite{Price},
the initial analysis was again at 1 km from the cascade in the plane of the pancake. 
(The pancake is at a depth about 5\% further from the shower origin due to the relatively lower density of Antarctic ice to that of sea water). 
In figure~\ref{fig8}a) the detected acoustic pulse is plotted for three energies. The pulse shape again scales with energy but is 5-6 pPa 
per GeV, (about 5 times that of water). The pulse is narrower than in water, though not by the factor of 2.6 as predicted by the 
relative velocities. The pulse is more symmetric as the attenuation is dominated almost entirely by scattering and is non-complex. 
The pulse also shows a ripple indicative of the sharp $f^4$ nature of Rayleigh scattering. In figure~\ref{fig8}b) the relationship between the 
maximum pulse height and energy is displayed. Again the pulse height grows almost linearly with energy and the best straight line fit 
of $\log_{10}p$ to $\log_{10}E$ as in equation 20 yields a slope of 1.0079 ($\pm$ 0.0017)  and an intercept of -11.349 ($\pm$ 0.151), corresponding again to a growth in 
pulse height which is very slightly greater than energy deposition and an intercept corresponding to 4.48~pPa/GeV. In practice it is 
the higher energies which are of most interest. If the fit is done in the 10$^9$-10$^{12}$~GeV region the intercept corresponds to 5.35~pPa/GeV. 
The average frequency is nearly constant at 39~kHz and the asymmetry is nearly zero.

\begin{figure}
\includegraphics[width=\textwidth]{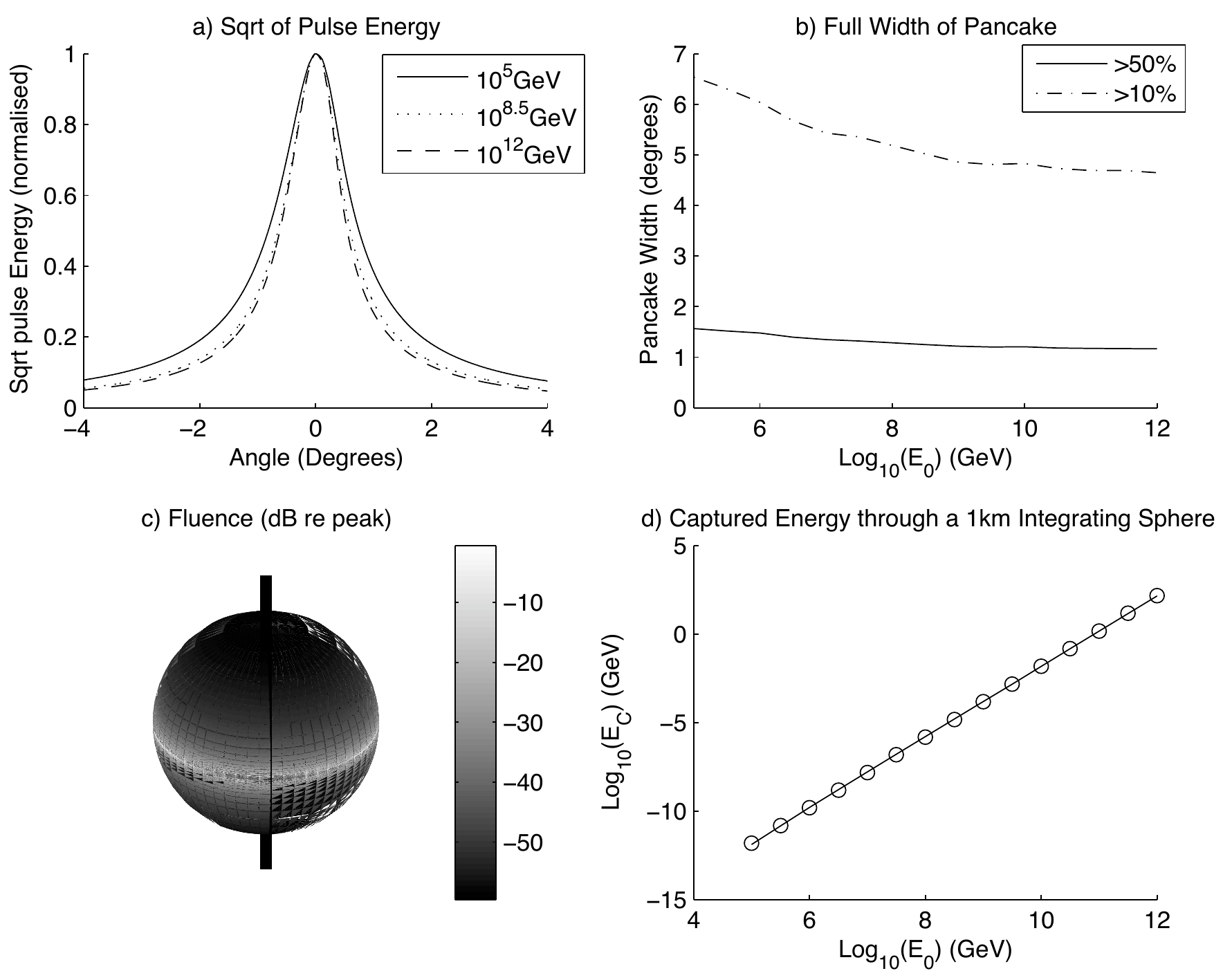}
\caption{Variation of acoustic pulse energy with angle in ice
a) square root of the pulse energy versus angle
b) full width for 10\% and 50\% of maximum versus energy
c) fluence through an integrating sphere
d) captured energy versus deposited energy and linear fit (see equation 22).}
\label{fig9}
\end{figure}

Consider now figure~\ref{fig9} showing the angular spread of the pancake in ice. The analysis is identical to that described in
section 4.3. As can be seen in figures~\ref{fig9}a) and \ref{fig9}b), the angular spread of the pulse is very similar to that of sea water, 
but is slightly broader due to the effective wavelength in ice being longer than that of water. In figures~\ref{fig9}c) and \ref{fig9}d) the energy captured 
on an integrating sphere at 1 km is calculated. In figure~\ref{fig9}d) the scaling of captured energy with the square of the deposited energy is again 
evident. The intercepts are -22.19~GeV and -21.76~GeV with and without attenuation respectively. Hence about 11 times as much energy 
is created in ice as in water. Once attenuation is included the energy drops to 37\%; 63\% of the energy is lost nearly entirely due 
to scattering. This loss is slightly more than water largely because ice pulses are at a higher frequency causing more effective attenuation.
 
\begin{figure}
\includegraphics[width=\textwidth]{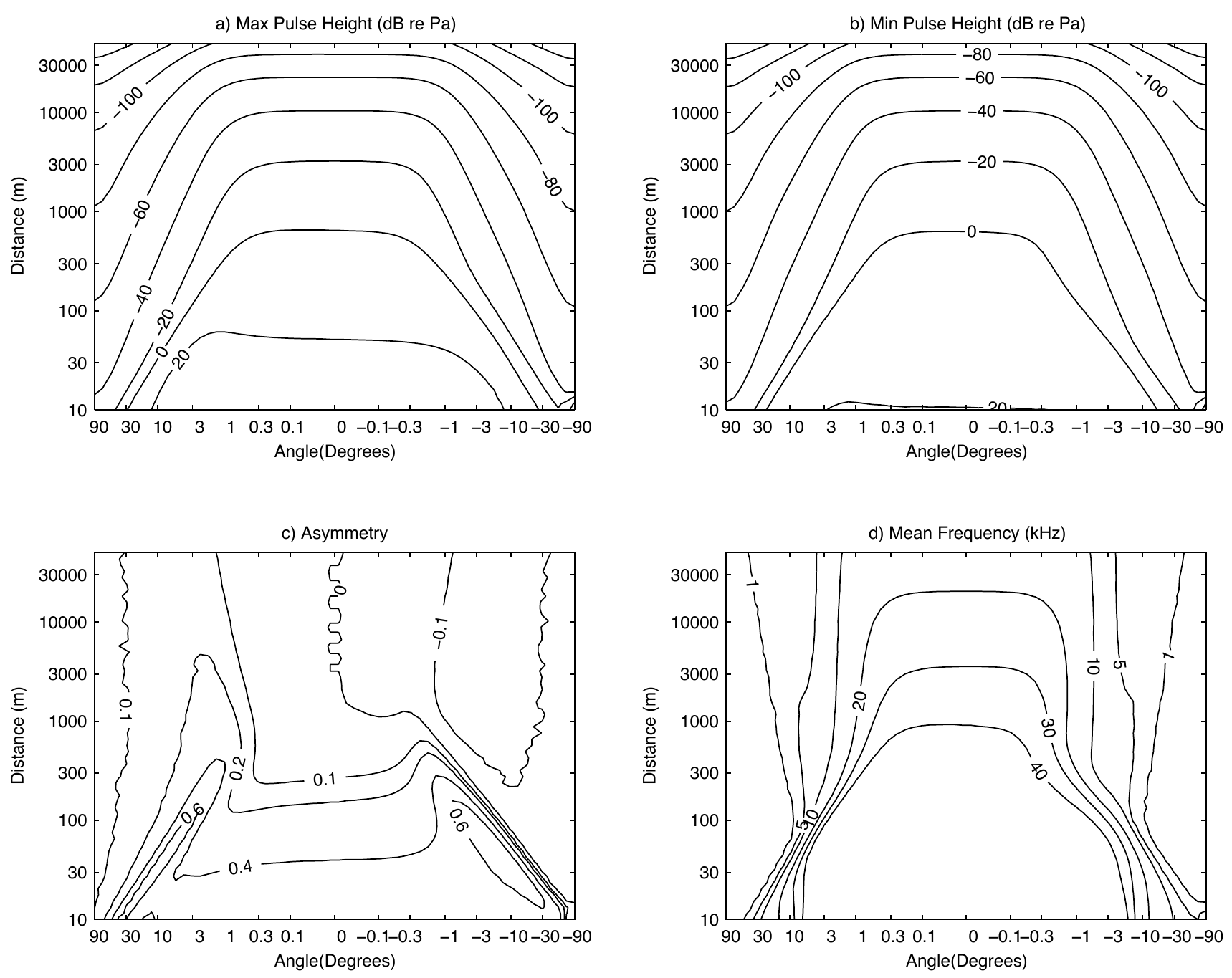}
\caption{Pulse parameters in ice for a $10^{11}$GeV primary as a function of distance and angle
a) maximum pulse height
b) minimum pulse height
c) pulse asymmetry
d) mean frequency.}
\label{fig10}
\end{figure}

As with sea water the analysis is extended to distances between 10~m and 50~km and angles between $\pm90^\circ$. The mean frequency, 
maximum and minimum pulse heights and asymmetry are displayed in figure~\ref{fig10}. Here the mean frequency is defined as the 
frequency of a single cycle sine wave that most closely represents the acoustic bipolar pulse, this is usually a little higher than
the peak frequency.
As anticipated both the frequencies and pulse height 
are greater than that of sea water. Also, due to the real nature of the attenuation the asymmetry falls rapidly to zero in the plane of 
the pancake and the odd symmetry in the far field between plus and minus angles is also more evident.
\section{Modelling showers and parameterising the fluctuations}
\subsection{Introduction}
Previous work in this area has \cite{ACORNE}\cite{NiessBertin}\cite{SAUND}
concentrated on modelling average shower parameters as a function of energy. Fluctuations have not been 
considered. Indeed due to the nature of previous parameterisations which involve correlated variables, the inclusion of fluctuations is not feasible as varying 
one parameter will mean that the other parameters have also to be tuned. The strategy in this section is to parameterise the shower and its fluctuations 
using an orthogonal basis set.
\subsection{CORSIKA Monte Carlo data}
\begin{figure}
\includegraphics[width=\textwidth]{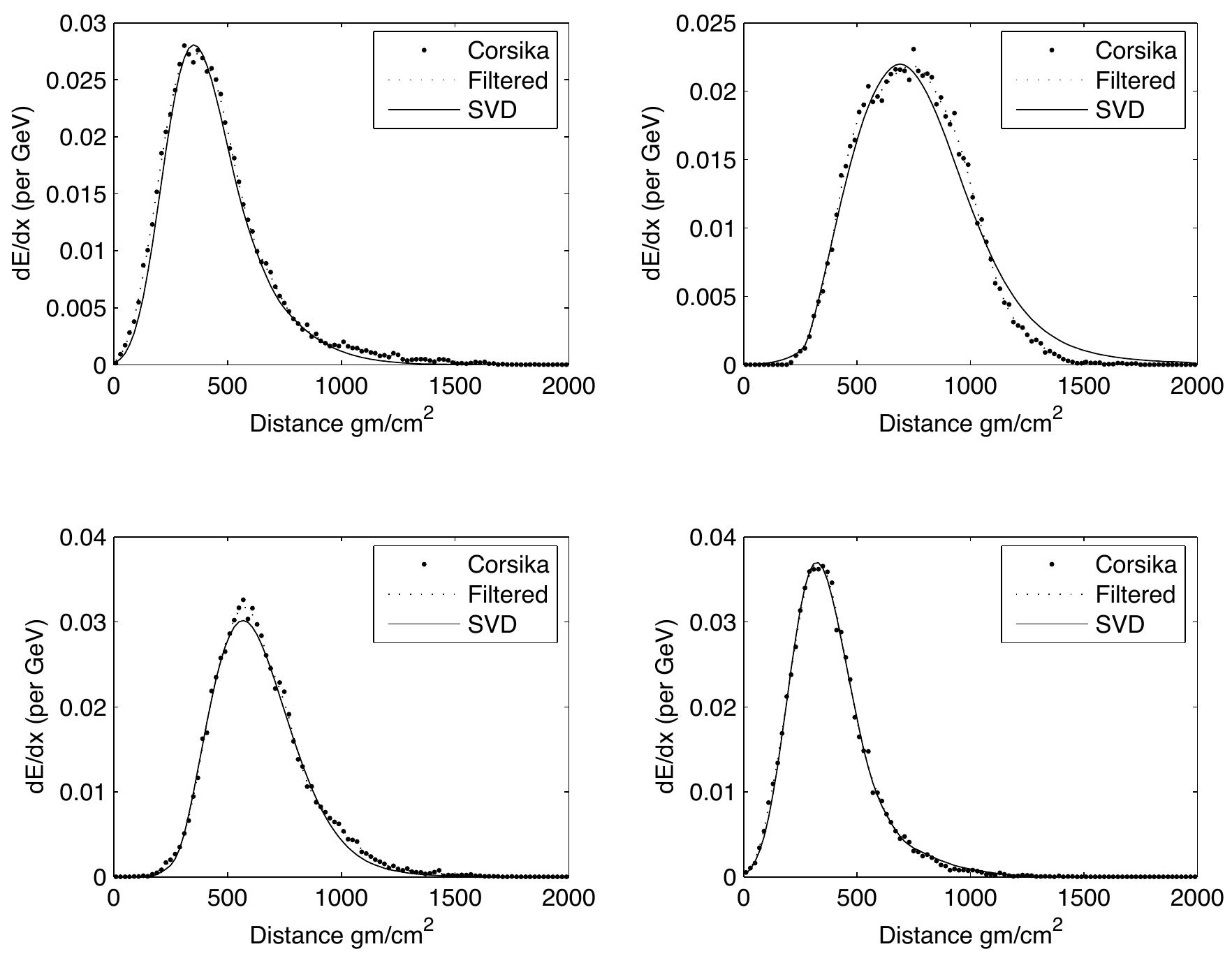}
\caption{Longitudinal distribution of four CORSIKA generated $10^5$ GeV showers (points) with smoothing (dotted curve) and SVD reconstruction
(solid curve).}
\label{fig11}
\end{figure}

The showers have been modelled with CORSIKA \cite{CORSIKA}, which uses a thinning process. 
The stochastic fluctuations in individual showers were smoothed using a non-causal 3rd order Butterworth filter.
In figure~\ref{fig11} the effect of the Butterworth filtering is 
shown for four typical $10^5$ GeV showers. The overall shape of the shower is retained but the noise considerably reduced.
\subsection{ SVD parameterisation}
%
%
Parameterisation using singular value decomposition (SVD) \cite{SVD} is an eigenvector based technique and has a number of advantages
which include applicability to data for which functional parameterisation is difficult and, most importantly, the ability to include 
fluctuations. The ability to include fluctuations stems from the orthogonal nature of the parameterisation; each parameter can be varied 
independently as there is no covariance.
The standard SVD method is matrix based and directly applicable in a 2d parameterisation. In the case of a 3d parameterisation the SVD has 
to be applied a number of times; we use two in this study. As the radial distribution is most critical for the acoustic pulse shape, 
this is used as our primary dimension. Each of our 1500 CORSIKA generated showers 
$\bf{S_k}$ is treated as a matrix with 100 rows corresponding to longitudinal distance and 20 columns corresponding to radius. These are appended to form: 
\begin{equation}
{\bf{O = }}\left( \begin{array}{l}
 {\bf{S}}_{\bf{1}}  \\ 
 {\bf{S}}_{\bf{2}}  \\ 
 {\bf{.}} \\ 
 {\bf{.}} \\ 
 {\bf{S}}_{\bf{k}}  \\ 
 \end{array} \right),
\end{equation}
creating an observation matrix \bf O \rm of size 150,000x20. A singular value decomposition was performed on the \bf O \rm{matrix}:
\begin{equation}
{\bf{O = WLV}} = \left( {{\bf{W}}_{\bf{S}} {\bf{W}}_{\bf{N}} } \right)\left( {\begin{array}{*{20}c}
   {{\bf{L}}_{\bf{S}} } & {\bf{0}}  \\
   {\bf{0}} & {{\bf{L}}_{\bf{N}} }  \\
\end{array}} \right)\left( {\begin{array}{*{20}c}
   {{\bf{V}}_{\bf{S}} }  \\
   {{\bf{V}}_{\bf{N}} }  \\
\end{array}} \right),
\end{equation}
where {\bf W} and {\bf V} are unitary containing the column and row eigenvectors of $\bf{OO^T}$ respectively sorted in 
decreasing magnitude of the associated eigenvalue and {\bf L} is diagonal containing the square roots of the respective eigenvalues. The $S$ and $N$ 
subscripts correspond to signal and noise and are discussed below. 
If all 20 eigenvectors are used the data can be reproduced perfectly. 
With fewer eigenvectors the data can be approximated with an accuracy  $\alpha$  given by:

\begin{equation}
\alpha=\frac{\sum\limits_{i = 1}^{i = n} {\bf{L}}_{ii}}{\sum\limits_{i = 1}^{i = 20} {{\bf{L}}_{ii} }}.
\end{equation}
The matrix has been partitioned into a signal space and noise space
by choosing the $n$ eigenvectors which contain the most information and assigning those to the signal space, the remaining
$20-n$ eigenvectors become the noise space:
\begin{equation}
{\bf{O = O}}_{\bf{S}} {\bf{ + O}}_{\bf{N}} {\bf{ = W}}_{\bf{S}} \left( {\begin{array}{*{20}c}
   {{\bf{L}}_{\bf{S}} }  \\
   {\bf{0}}  \\
\end{array}} \right){\bf{V}}_{\bf{S}} {\bf{ + W}}_{\bf{N}} \left( {\begin{array}{*{20}c}
   {\bf{0}}  \\
   {{\bf{L}}_{\bf{N}} }  \\
\end{array}} \right){\bf{V}}_{\bf{N}}. 
\end{equation}
%

The choice of the value of $n$ can be somewhat subjective, however choosing $n=4$ yields $\alpha=0.95$, which is more than sufficient given the other errors in the process.  
The matrix $\bf V_s$ (20x4) consists of the four row eigenvectors. 
The matrix $\bf W_S$  has four columns (4x150,000), this can be partitioned as follows: 
\begin{equation}
 \begin{array}{l}
   {\bf{W}}_{\bf{S}} {\bf{ = }}
   \left( 
   {\begin{array}{*{20}c}
    {{\bf{W}}_{{\bf{11}}} } & {{\bf{W}}_{{\bf{21}}} } & {{\bf{W}}_{{\bf{31}}} }  \\
    {{\bf{W}}_{{\bf{12}}} } & {{\bf{W}}_{{\bf{22}}} } & {{\bf{W}}_{{\bf{32}}} }  \\
    {\bf{.}} & {\bf{.}} & {\bf{.}}  \\
    {\bf{.}} & {\bf{.}} & {\bf{.}}  \\
    {{\bf{W}}_{{\bf{1k}}} } & {{\bf{W}}_{{\bf{2k}}} } & {{\bf{W}}_{{\bf{3k}}} }  \\
   \end{array}
   \begin{array}{*{20}c}
    {{\bf{W}}_{{\bf{41}}} }  \\
    {{\bf{W}}_{{\bf{42}}} }  \\
    {\bf{.}}  \\
    {\bf{.}}  \\
    {{\bf{W}}_{{\bf{4k}}} }  \\
   \end{array}}
   \right),
 \end{array}
\end{equation}
where each of the sub matrices $\bf{W_{mn}}$ is of size 100x1. However these have the shower information sequentially and need to be reshaped to look at the longitudinal correlations between showers. The matrix is reordered into four matrices of size 1500x100:
\begin{equation}
 \begin{array}{l}
   {\bf{W}}_{\bf{S}} {\bf{ = }}
   \left( 
    \begin{array}{*{20}c}
    {\bf{W}_{\bf{11}}^{\bf{T}} }  \\
    {\bf{W}_{\bf{12}}^{\bf{T}} }  \\
    {\bf{.}}  \\
    {\bf{.}}  \\
    {\bf{W}_{\bf{1k}}^{\bf{T}} }  \\
    \end{array}
   \right)
   ...
   \left(
   {\begin{array}{*{20}c}
    {\bf{W}_{\bf{41}}^{\bf{T}} }  \\
    {\bf{W}_{\bf{42}}^{\bf{T}} }  \\
    {\bf{.}}  \\
    {\bf{.}}  \\
    {\bf{W}_{\bf{4k}}^{\bf{T}} }  \\
   \end{array}}
   \right)
 \end{array}
\end{equation}
A further singular value decomposition is applied to each of these matrices in turn. As these matrices contain successively less information 
four of the column eigenvectors were retained for $\bf W_1$, three for $\bf W_2$, two for $\bf W_3$ and one for $\bf W_4$. 
The four $\bf L_m$ and $\bf V_m$ matrices of this second SVD contain the information yielding the 10 parameter values for each 
shower. These parameters may be recovered by creating a matrix:
\begin{equation}
{\bf{A = }}\left( {\begin{array}{*{20}c}
   {{\bf{V'}}_{\bf{1}} } & {{\bf{V'}}_{\bf{2}} } & {{\bf{V'}}_{\bf{3}} } & {{\bf{V'}}_{\bf{4}} }  \\
\end{array}} \right)\left( {\begin{array}{*{20}c}
   {{\bf{L'}}_{\bf{1}} } & 0 & 0 & 0  \\
   0 & {{\bf{L'}}_{\bf{2}} } & 0 & 0  \\
   0 & 0 & {{\bf{L'}}_{\bf{3}} } & 0  \\
   0 & 0 & 0 & {{\bf{L'}}_{\bf{4}} }  \\
\end{array}} \right).
\end{equation}
The matrix $\bf A$ (1500x10) contains on each row the coefficients $a_1$ to $a_{10}$ for each shower in turn.
The original showers, $\bf{S_k}$, can simply be reconstructed by matrix multiplication of the appropriate coefficients:
\begin{equation}
\begin{array}{l}
 {\bf{S}}_{\bf{k}}  = a_{k,1} {\bf{W}}_{{\bf{11}}} {\bf{V}}_{\bf{1}}  + .. + a_{k,4} {\bf{W}}_{{\bf{14}}} {\bf{V}}_{\bf{1}}  + a_{k,5} {\bf{W}}_{{\bf{21}}} {\bf{V}}_{\bf{2}}  + . \\ 
  + a_{k,7} {\bf{W}}_{{\bf{23}}} {\bf{V}}_{\bf{2}}  + a_{k,8} {\bf{W}}_{{\bf{31}}} {\bf{V}}_{\bf{3}}  + a_{k,9} {\bf{W}}_{{\bf{32}}} {\bf{V}}_{\bf{3}}  + a_{k,10} {\bf{W}}_{{\bf{41}}} {\bf{V}}_4.  \\ 
 \end{array}
\end{equation}
The showers are recreated to within an average accuracy of 5 \% in profile and magnitude.
Lower energy showers have the greatest variation in shape. In figure~\ref{fig11} 
the longitudinal distribution of four of the most extreme showers are plotted. These are chosen by having the greatest and least component 
of the first two radial eigenvectors. These will be some of the most difficult showers to model. The reconstructed showers though clearly 
not perfect do however reasonably fit the data.
The mean and standard deviation of the parameters in each of the 15 half decades can now be used to reproduce the statistics of the showers, $a_1$ to $a_{10}$. 
Interpolation can be used to model intermediate values. Care must be taken however as these are not orthogonal as is normally the case with the SVD. 
Because of the linear nature of the parameterisation however these correlations can be reintroduced by multiplying by the matrix square root of the 
correlation matrix: 
\begin{equation}
{\bf{A}}(E) = {\bf{\bar A}}(E) + \left( {\sum\limits_{i = 1}^{10} {{\bf{R}}_i {\bf{A}}_\sigma  (E)_i } } \right){\bf{C}}(E)^{\frac{1}{2}}, 
\end{equation}
where ${\bf \bar A}(E)$ is a vector containing the average $a_i$ values at a given energy, ${\bf R}$ is a vector of normally distributed random 
numbers with a mean of 0 and a standard deviation of 1, ${\bf A}_{\sigma} (E)$ are the standard deviations of $a_i$ and ${\bf C}(E)$ is the 
matrix of the correlation coefficients between parameters within a shower and averaged over all showers at a particular energy.
The parameter values are too cumbersome to list here 
but are available on the ACORNE web site \cite{ACORNEURL}.
\subsection{Fluctuations applied to acoustic pulses}
\begin{figure}
\includegraphics[width=\textwidth]{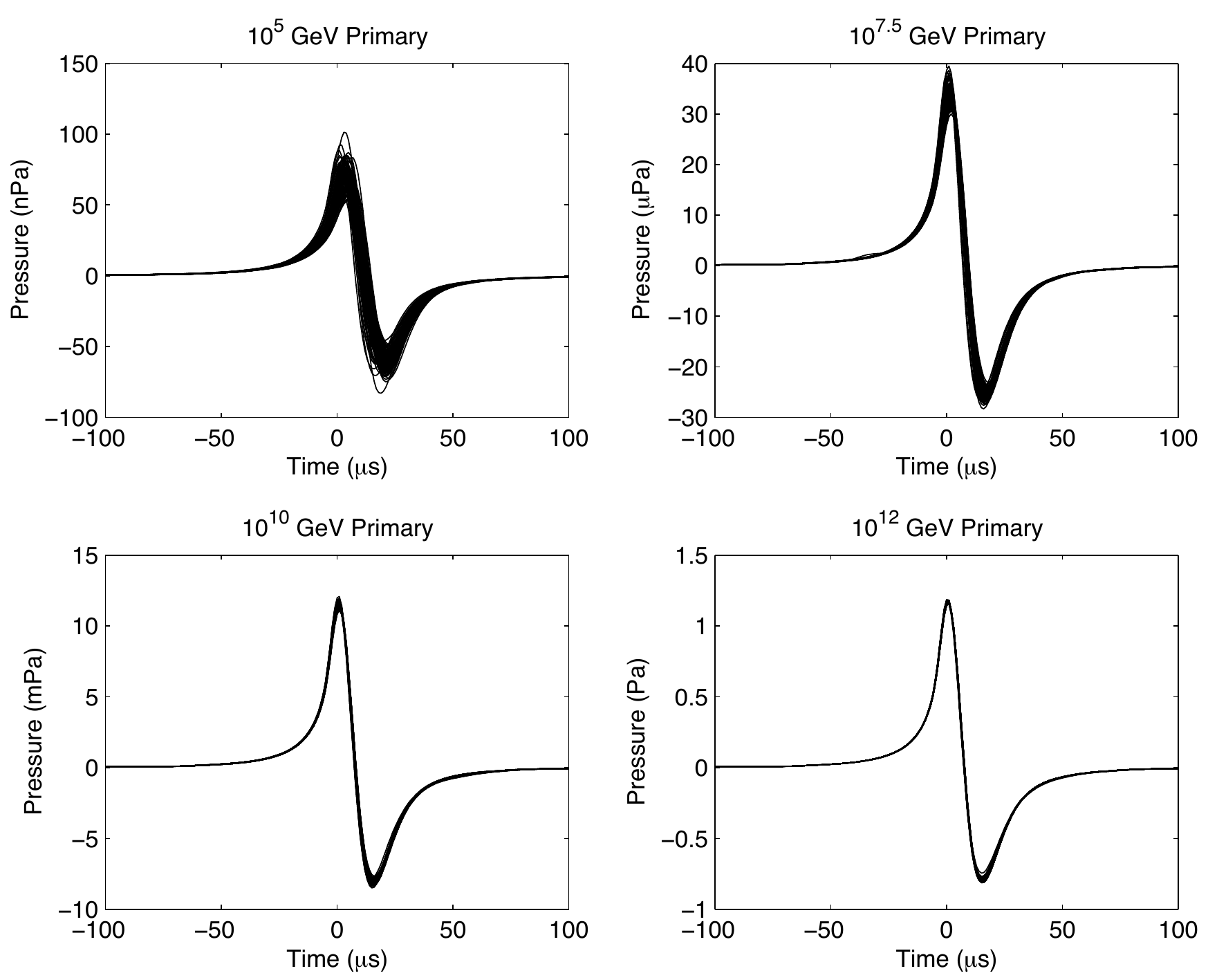}
\caption{Fluctuation of acoustic pulses at 1 km from the shower for $10^5$, $10^{7.5}$, $10^{10}$ and $10^{12}$ GeV. }
\label{fig12}
\end{figure}

\begin{figure}
\includegraphics[width=\textwidth]{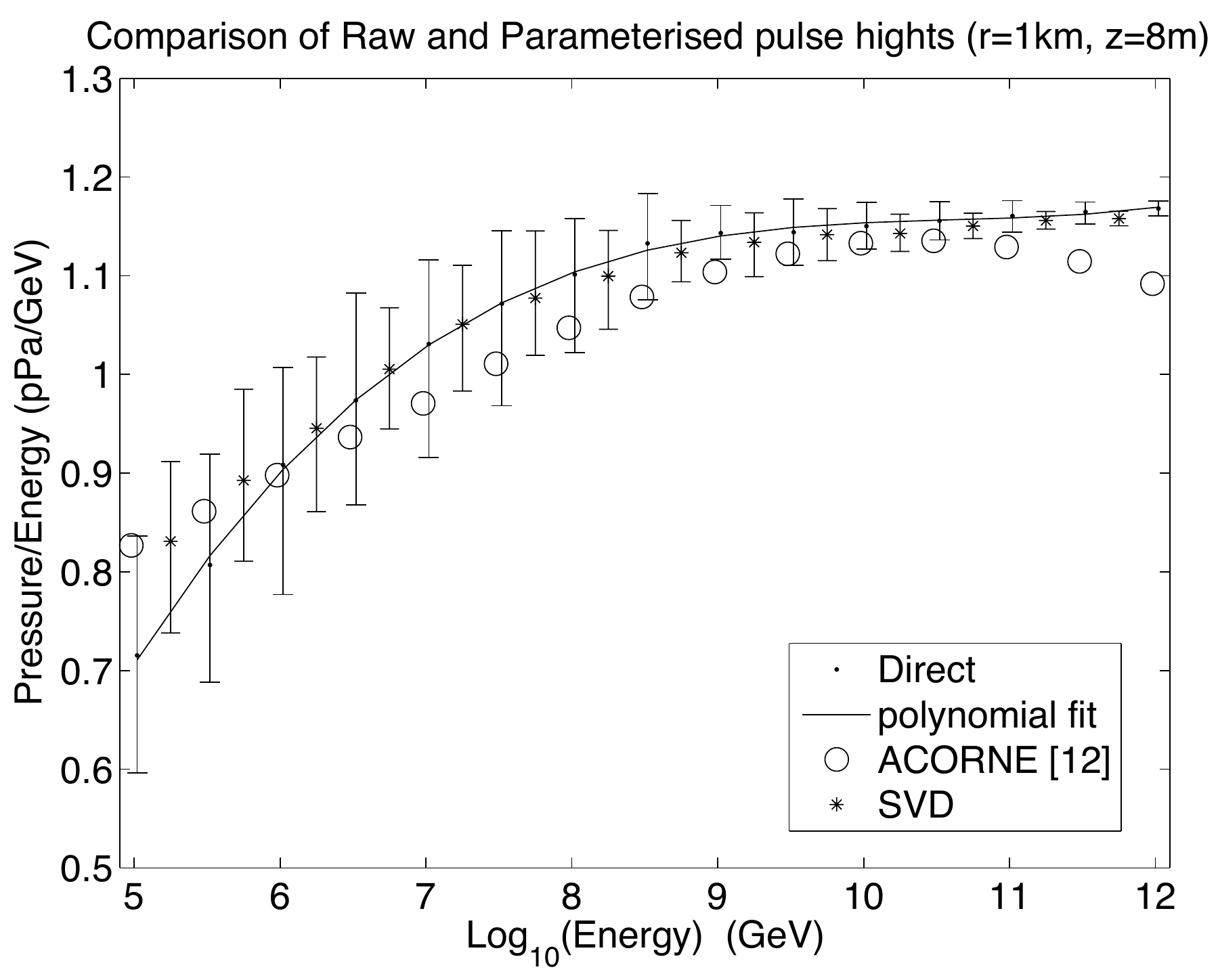}
\caption{Pulse heights for CORSIKA generated showers at 1km from the source as a function of energy, 
with the SVD parameterisation and functional parameterisation shown for comparison.
The 10th and 90th centiles are shown for the CORSIKA and SVD cases.}
\label{fig13}
\end{figure}

In figure~\ref{fig12} the anticipated pulses from 100 CORSIKA generated events for each of the energies $10^5$, $10^{7.5}$, $10^{10}$ and $10^{12}$ GeV are plotted. 
These show clear evidence that fluctuations become less important with increasing energy.
In figure~\ref{fig13} the maximum acoustic pulse height per GeV at $z$ =8 m and $r$=1 km from the shower is presented. For both the original 
and SVD parameterised data one hundred showers are modelled in each half decade. The output directly from Corsika is used to create the ``direct'' plot. 
The pulse height from interpolated SVD parameterisation are plotted at 1/4 and 3/4  decades for clarity. The error-bars are drawn at the 10 and 90 centiles. 
The earlier ACORNE parameterisation \cite{ACORNE} is shown for comparison. Neither of the parameterisations is perfect, however the differences are small; 
only a few percent at all energies. The SVD parameterisation model seems to slightly under-estimate the fluctuations. This is due to the statistics being 
slightly non Gaussian (leptokurtic). The earlier ACORNE parameterisation also works well and produces pulses with similar accuracies.
\subsection{Summary of the SVD parameterisation}
The SVD technique successfully models the shape of both the radial and longitudinal distributions of the showers to a high 
accuracy ($>95\%$ of the information is retained). 
However for the acoustic detection of cosmic ray neutrinos it is primarily energies of greater than 10$^9$GeV that are of
significance to the acoustic technique. 
In this region fluctuations only contribute a few percent to the acoustic pulse amplitude and our earlier parameterisation \cite{ACORNE} 
is perfectly adequate for modelling pulses in this region.
In broader terms however the SVD method outlined is a linear technique and does not rely on optimisation and as the accuracy is dictated 
by the number of eigenvectors, a trade off can be made between accuracy and complexity.
\section{Conclusions}
A new way of computing the acoustic signal for neutrino showers in water and
ice has been described. The method is computationally fast and allows the
most up to date knowledge of the attenuation to be incorporated naturally.
This is now known to be complex in nature. A parameterisation of this
attenuation is given. The properties of the expected acoustic signals from
such showers have been described. These properties will be used in a future
search for such interactions in an acoustic array. A matrix method of
parameterising the signals which includes fluctuations has also been
described. This method is based on the SVD technique and it is shown to
model both the shape of the radial and longitudinal distributions of the
showers to a good accuracy.
\newpage
\end{document}